# Wavelength-Selective Switches for Mode-Division Multiplexing: Scaling and Performance Analysis

Keang-Po Ho, *Senior Member, IEEE,* Joseph M. Kahn, *Fellow, IEEE,* Jeffrey P. Wilde, *Member, IEEE*

*Abstract*—Wavelength-selective switches for mode-division-multiplexing systems are designed by scaling switches from single-mode systems. All modes at a given wavelength are switched as a unit, which is necessary in systems with substantial mode coupling, and minimizes the number of ports required to accommodate a given traffic volume. When a pure mode is present at the input, modal transmission and coupling coefficients are mode-dependent and may be computed using a simple mode-clipping model. When multiple modes are present, interference between modes alters the transmission and coupling coefficients, shifting the passband center frequency and changing its bandwidth. Mode-coupling matrices are used to compute mixed modes having the narrowest or widest bandwidths, or having the largest center-frequency offsets. In a specific design for graded-index fiber, five mode groups and 50-GHz channel spacing, the one-sided bandwidth may change up to $\pm 3.6$ GHz. In a system with many cascaded switches and strong mode coupling, the end-to-end response per switch may be characterized by a mode-averaged transmission coefficient.

*Index Terms*—Wavelength-selective switch, multimode fiber, mode-division multiplexing.

## I. Introduction

IN mode-division-multiplexed (MDM) systems, multiple data streams are transmitted in different modes of multimode fiber (MMF) [1]-[7]. Ideally, transmission capacity increases in proportion to the number of modes [6][7]. In addition to spatial multiplexing, MDM systems use wavelength-division multiplexing (WDM) to fully utilize the bandwidth available in the MMF and inline optical amplifiers. Reconfigurable optical add-drop multiplexers (ROADMs) [9]-[12] are indispensable for dynamically reconfigurable optical networks. To ensure the viability of MDM in such systems, ROADMs for MDM should achieve functionality and performance similar to their counterparts in single-mode fiber (SMF) systems.

In ROADMs for long-haul MDM systems, it is desirable to switch all the modes at a given wavelength as a unit between the same input and output ports [13]-[20]. In all long-haul MDM systems to date, mode coupling occurring along the link has been compensated by joint multi-input multi-output (MIMO) signal processing of all modes at the receiver [1]-[7], which requires all modes to be switched as a unit. Moreover, switching all modes as a unit simplifies network management and minimizes the number of ROADM input and output ports required to accommodate a given aggregate traffic volume [15].

Wavelength-selective switches (WSSs) are a principal component in ROADMs [9]-[12]. For implementation of the switching plane in a WSS, liquid-crystal-on-silicon (LCoS)-based spatial light modulators (SLMs) [21]-[24] offer several advantages over previous technologies, and have become increasingly popular in recent years.

The complex modal profiles of the signals in MMF are the main complication in the design of a multimode WSS. The field distribution at an input port is a speckle pattern determined by the combination of modes launched into the MMF and by mode coupling during propagation through the MMF [7][25]. In single-mode fiber, by contrast, regardless of the launched field profile, after propagating just a few meters, the output field profile is always the same [26].

This paper addresses the design and performance of LCoS-based multimode WSSs. Starting with a single-mode WSS, certain physical dimensions within the WSS are scaled with the goal of accommodating multiple modes while maintaining performance objectives, such as isolation, insertion loss, bandwidth, passband ripple, and passband symmetry. Methods to analyze the mode-dependent transmission response of multimode WSSs are developed and applied. Pure modes with different mode sizes are subject to variations in passband shape and bandwidth that are consistent with a simple mode-clipping model. Pure modes also become coupled to each other, especially at frequencies near the passband edge. A matrix describing this mode coupling can be used to determine the mixed modes having the narrowest or widest bandwidths. Another matrix can be used to find the mixed modes having maximum center-frequency offset.

The remainder of this paper is organized as follows. Section

Manuscript received January ??, 2014, revised April ??, 2014. This research of JMK was supported in part by National Science Foundation Grant Number ECCS-1101905 and by Corning, Inc.

K.-P. Ho is with Silicon Image, Sunnyvale, CA 94085 (Tel: +1-408-419-2023, Fax: +1-408-616-6399, e-mail: kpho@ieee.org).

J. M. Kahn and J. P. Wilde are with E. L. Ginzton Laboratory, Department of Electrical Engineering, Stanford University, Stanford, CA 94305, USA (e-mail: jmk@ee.stanford.edu, jpwilde@stanford.edu).

II describes how to scale a WSS from single- to multi-mode operation based on the beam spot size. Section III compares passband responses of multimode WSSs obtained by detailed simulation to those computed using a simple mode-clipping model, and derives a mode-coupling matrix using the mode-clipping model. Section IV uses the mode-clipping model to analyze wavelength-selective filtering of mixed modes, discusses mode-averaged filtering in long-haul systems with strong mode coupling, and presents a method to determine the mode mixtures having the narrowest or widest bandwidths, or the worst center-frequency offsets. Section V discusses the impact of filtering on optical signals. Section VI concludes the paper.

## II. WSS Scaling for Multimode Operation

Figure 1 shows a simplified schematic of an LCoS-based WSS, and is applicable to single- or multi-mode devices. The input/output ports comprise a linear array of fibers with collimating lenses. The labeling of input and output ports in Fig. 1 assumes a drop module. A ruled grating between the collimating lenses and a Fourier lens maps signals at a given wavelength to/from the appropriate switching segment on the LCoS SLM, independent of the input and output ports.

In Fig. 1, the vertical centerlines of the ruled grating and the LCoS SLM are assumed to lie in the two focal planes of the Fourier lens, thereby making the system telecentric at the SLM plane (i.e., the chief ray associated with any port is normal to the SLM plane for all wavelengths in the range of device operation). Applying a linear phase ramp along the beam-steering direction ($y$-axis) switches a signal between different output ports. The system essentially images a fiber output onto the SLM with a magnification along the $y$-axis given by the ratio of the Fourier lens focal length to the collimator lens focal length and along the $x$-axis by this same ratio times a factor associated with the anamorphic scaling of the beam by the grating. Polarization-diversity and additional anamorphic beam-transformation optics are not shown in Fig. 1 for simplicity.

In this section, the WSS of Fig. 1 is analyzed taking account of the increased spot size of a multimode beam compared to a single-mode beam. This analysis yields simple scaling relationships from single- to multi-mode WSSs.

### A. Laguerre-Gaussian Modes

We consider graded-index MMF, which has far lower group-delay spread than step-index MMF (assuming more than two mode groups), which is important for minimizing receiver MIMO signal processing complexity [7][27][28]. Although a practical MMF has a finite core radius to support a finite number of modes, for analytical convenience, we consider the eigenmodes of an infinite parabolic index profile, which are given in polar coordinates $(\rho,\phi)$ by the Laguerre-Gaussian (LG) functions:

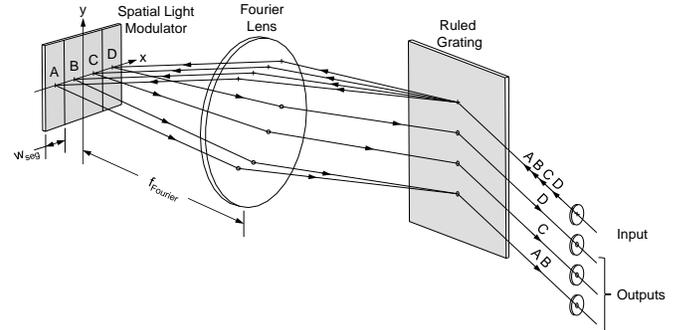

Fig. 1. Schematic design of a WSS (drop module) using an LCoS SLM to switch input signals between output ports. Beam-transformation and polarization-diversity optics are not shown for simplicity.

$$E_{q,m}(\rho,\phi) = C_{q,m} \frac{\rho^m}{w_0^{m+1}} L_q^{(m)}\left(\frac{2\rho^2}{w_0^2}\right) \exp\left(-\frac{\rho^2}{w_0^2}\right) \begin{Bmatrix} \sin m\phi \\ \cos m\phi \end{Bmatrix}. \quad (1)$$

The indices $q$ and $m$ are the radial and azimuthal orders, respectively, $C_{q,m} = \left[2^{m+1}(2-\delta_{m,0})q!/\pi(q+m)!\right]^{1/2}$ is a parameter normalizing the mode to unit energy, $\delta_{m,n}$ is the Kronecker delta, equal to 1 only if $m = n$, $L_q^{(m)}(\cdot)$ is a generalized Laguerre polynomial, and $w_0$ is the $1/e$ radius of the fundamental LG$_{00}$ modal field [setting $q = m = 0$ in (1)]. The sine and cosine modes are defined here such that $\phi = 0$ coincides with the SLM frequency-spreading direction ($x$-axis) in Fig. 1.

In a parabolic-index MMF, all modes with a given value of $g = 2q + m + 1$ form a group having similar propagation constants. Although the Hermite-Gaussian modes [29]-[31] can describe the eigenmodes of a parabolic-index MMF in Cartesian coordinates, the LG modes are easier to separate into groups with similar propagation constants. For a given $(q,m)$, with nonzero azimuthal order $m \neq 0$, sine and cosine modes represent two degenerate modes with identical propagation constants. Including two polarizations, for a given $(q,m)$ pair, there are two modes for $m = 0$ and four modes for $m \neq 0$. The total number of propagating mode groups is denoted by $g_{\max}$, and the total number of propagating modes in two polarizations is denoted by $D$, where $g_{\max} = 1,2,3,4,5,...$ corresponds to $D = 2,6,12,20,30,...$. The Fourier transform of an LG mode is a scaled version of the same LG mode, a property that simplifies the analysis for a system using a Fourier lens, as in Fig. 1.

When an LG mode propagates from the input to the output of Fig. 1, the radius $w_0$ is a function of propagation distance $z$, $w_0(z)$. In other words, the LG modes change in size and phase profile as they propagate. In later parts of this paper, our notation does not make explicit this $z$-dependent mode radius, since we always compare mode sizes in equivalent planes, e.g. at the SLM surface or in the collimating lens plane.

### B. Mode Size Scaling

The spatial extent of LG modes generally increases with the group number $g = 2q + m + 1$. Figure 2 shows the intensity profiles of several LG-cosine modes. Figure 2 shows that the $LG_{04}$ and $LG_{12}$ modes, both in the $g = 5$ group with $m \neq 0$, have larger size than the $LG_{00}$ mode. The root-mean-square (RMS) radius of a mode is the square-root of $\int_{-\pi}^{+\pi}\int_{0}^{\infty}\rho^2 E_{q,m}^2 \rho d\rho d\phi$, and is equal to $\sqrt{g/2}\,w_0$.

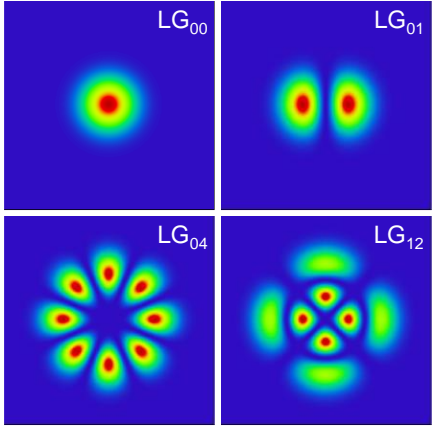

Fig. 2. Intensity profiles of selected LG-cosine modes.

In this section, we discuss scaling of certain dimensions within a WSS to accommodate the increased spatial extent of a multimode beam. We take the fundamental mode radius $w_0$ as given, and study how the beam size scales with the number of modes $D$. Given a value of $w_0$, we quantify the increased spatial extent of a multimode beam by a scale factor $\kappa$. We compute the scale factor $\kappa$ based on three different criteria, which are the numerical aperture (NA), or the radius containing 95% or 99% of the beam energy.

In experimental characterization, a fiber's NA[1] is defined as the sine of a half-angle spanning the far-field beam from its peak intensity to 5% of peak intensity [32]-[34]. For SMF, the NA is defined unambiguously by the half-angle at 5% of the peak intensity. In MMF, the NA in general depends on the mixture of modes excited. Here, the NA is defined by the sine of the half-angle at 5% of peak intensity measured with an over-filled launch that excites all propagating modes with equal power. In Fig. 3(a), the left axis shows the NAs of MMF supporting different numbers of modes $D$, normalized to the NA of the fundamental $LG_{00}$ mode. These normalized NAs are denoted by $\kappa_{NA}$. The right axis in Fig. 3(a) shows the corresponding effective beam radii at 5% of the peak intensity that define the fiber NA. To be consistent with notation below, these are denoted by $R_{eff}$. For the fundamental $LG_{00}$ mode, $R_{eff,0} = 1.22 w_0$. When referring to the fundamental mode, we denote the effective beam radius as $R_{eff,0}$.

It is important to note that the increase of the normalized NA $\kappa_{NA}$ with the number of modes does not imply that the NA itself increases. The NA for the fundamental mode may decrease with an increasing number of modes, as explained later.

---

[1] From geometric optics, a fiber's NA is given by $\sqrt{n_{core}^2 - n_{clad}^2}$, where $n_{core}$ and $n_{clad}$ are the core and cladding refractive indices, a definition that is independent of the core diameter or the number of propagating modes. The NA defined in this way deviates significantly from the NA as defined here when the number of propagating modes is small.

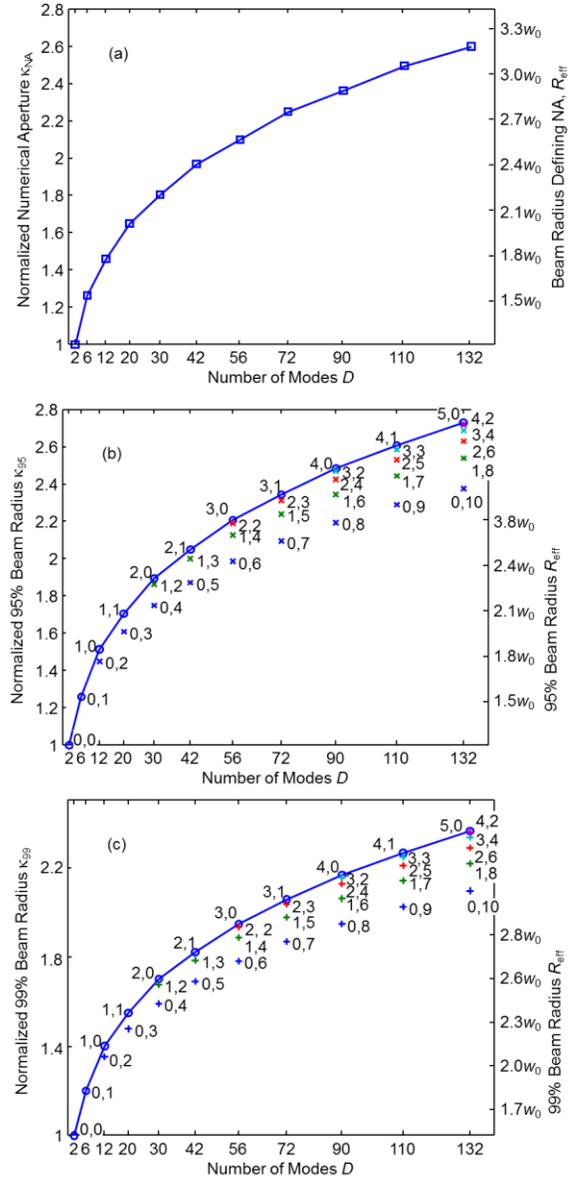

Fig. 3. Options for scale factor $\kappa$ as a function of the total number of modes $D$: (a) numerical aperture for overfilled launch, (b) 95% beam radius for pure modes, (c) 99% beam radius for pure modes (all three are normalized to the corresponding quantities for the fundamental mode). The right axes show the effective beam radii defining the scale factors. In (b) and (c), the symbols show the radius for individual LG modes, while the curves show the largest radius.

As an alternative to the NA, we consider scaling a WSS based on the effective beam radius enclosing 95% or 99% of the beam energy of the worst-case (largest) pure mode. For a MMF including mode groups up to $g_{max}$, numerical results show that the largest mode is typically the $LG_{(g_{max}-1)/2,0}$ mode for odd $g_{max}$ and the $LG_{g_{max}/2-1,1}$ mode for even $g_{max}$. In Figs. 3(b) and (c), the left axes show the effective beam radii $R_{eff}$ enclosing 95% or 99% of the beam energy for pure modes, normalized to the corresponding radii for the fundamental $LG_{00}$ mode. These normalized beam radii are denoted by $\kappa_{95}$ and $\kappa_{99}$, respectively. The right axes in Figs. 3(b) and (c) show the effective beam radii $R_{eff}$. For the fundamental $LG_{00}$ mode, the radii containing 95% and 99% of the energy are $R_{\text{eff},0} = 1.22w_0$ and $R_{\text{eff},0} = 1.52w_0$, respectively.

TABLE I. SCALING FROM SINGLE- TO MULTI-MODE WSS. HEAVY LINES DENOTE PARAMETERS DEFINING DESIGNS I, II AND III.
THE PARAMETERS $\kappa$ AND $\eta$ ARE DEFINED IN SECTION II.

| Component | Parameter | Design I | Design II | Design III | Design IV |
|---|---|---|---|---|---|
| Input/output fibers | Effective beam radius $R_{\text{eff}}$ relative to $R_{\text{eff},0}$ | $\kappa$ | $\kappa$ | $\kappa$ | $\kappa$ |
| | Fundamental mode radius $w_0$ or $R_{\text{eff},0}$ | $\eta$ | $\eta$ | $\eta$ | $\eta$ |
| LCoS SLM | Segment width $w_{\text{seg}}$ relative to $R_{\text{eff},0}$ | $\kappa$ | $\kappa$ | $\kappa$ | $\kappa$ |
| | Image eccentricity | 1 | 1 | 1 | 1 |
| | Pixel pitch | **1** | **$1/\kappa$** | **1** | **$1/\kappa$** |
| | Image of fundamental mode radius $w_0$ | 1 | $1/\kappa$ | 1 | 1 |
| | Segment width $w_{\text{seg}}$ in frequency direction $x$ | $\kappa$ | 1 | $\kappa$ | $\kappa$ |
| | Height in beam-steering direction $y$ | $\kappa$ | 1 | $\kappa$ | $\kappa$ |
| | Number of pixels in beam-steering direction $y$ | $\kappa$ | $\kappa$ | $\kappa$ | $\kappa^2$ |
| Ruled grating | Angular dispersion $\partial\theta/\partial\nu$ | **1** | **1** | **$\kappa$** | **1** |
| | Image of fundamental mode radius $w_0$ | $\kappa$ | $\kappa$ | 1 | $\kappa$ |
| | Overall dimensions in both directions | $\kappa^2$ | $\kappa^2$ | $\kappa$ | $\kappa^2$ |
| Fourier lens | Focal length $f_{\text{Fourier}}$ | $\kappa$ | 1 | 1 | $\kappa$ |
| | Radius | $\kappa^2$ | $\kappa^2$ | $\kappa$ | $\kappa^2$ |
| | $f$-number | $1/\kappa$ | $1/\kappa^2$ | $1/\kappa$ | $1/\kappa$ |
| Collimator lenses | Focal length $f_{\text{coll}}$ | $\eta\kappa$ | $\eta\kappa$ | $\eta$ | $\eta\kappa$ |
| | Radius | $\kappa^2$ | $\kappa^2$ | $\kappa$ | $\kappa^2$ |
| | $f$-number | $\eta/\kappa$ | $\eta/\kappa$ | $\eta/\kappa$ | $\eta/\kappa$ |
| Ports | Port spacing | $\kappa^2$ | $\kappa^2$ | $\kappa$ | $\kappa^2$ |
| | Port angular separation | $\kappa$ | $\kappa^2$ | $\kappa$ | $\kappa$ |
| | Number of ports | $1/\kappa$ | $1/\kappa^2$ | $1/\kappa$ | 1 |

### C. Wavelength-Selective Switch Scaling

In this subsection, we discuss how to scale the WSS of Fig. 1 from single- to multi-mode operation. In a multimode WSS, the beam radius is generally larger than that in a single-mode WSS, and increases with the number of modes $D$, as shown in Fig. 3. To accommodate the larger beam radius, the optical system must be modified if similar passband performance is to be maintained. Various options for modifying the design exist, but in all cases, specific optical components are scaled by factors related to $\kappa$. As discussed in Sec. II.B, various definitions for $\kappa$ exist, and for the remainder of the discussion, it is assumed that one has been chosen according to some design criteria and is simply referred to as $\kappa$ without a subscript. Like the fundamental mode radius $w_0(z)$, the effective beam radius $R_{\text{eff}}(z)$ changes with propagation distance $z$. This $z$-dependence is suppressed below, since mode sizes are always compared in equivalent planes.

Table I summarizes the scaling of key WSS component parameters for four different design approaches. Design I assumes that both the LCoS SLM pixel pitch and the ruled grating angular dispersion remain unchanged in scaling from single- to multi-mode operation. Design II scales the SLM pixel pitch, Design III scales the grating angular dispersion, and Design IV combines Designs I and II. These four designs are illustrative, and other designs are obviously possible. We first discuss Design I in detail, and then discuss the other designs.

The filtering performance of the WSS is determined by the ruled grating angular dispersion and the per-channel segment width of the SLM. In the WSS shown in Fig. 1, when two signals separated by the nominal channel spacing $\Delta\nu$ are input to one port, the corresponding rays are separated by an angle $(\partial\theta/\partial\nu)\Delta\nu$, where $\partial\theta/\partial\nu$ is the angular dispersion of the ruled grating. On the SLM, located in the focal plane of the Fourier lens with focal length $f_{\text{Fourier}}$, the corresponding image centroids are separated along the frequency direction ($x$-axis) by [35][36]

$$w_{\text{seg}} = \frac{\partial\theta}{\partial\nu}f_{\text{Fourier}}\Delta\nu. \quad (2)$$

The SLM is nominally subdivided into switching segments of width $w_{\text{seg}}$.

In a single-mode WSS, filtering performance is determined substantially by the switching segment width $w_{\text{seg}}$ relative to the image size of the fundamental mode radius $w_0$ (or equivalently the effective beam radius $R_{\text{eff},0}$). High isolation between adjacent WDM channels requires that their images on the SLM lie on disjoint switching segments.

Similarly, in a multimode WSS, filtering performance is determined by the ratio of $w_{\text{seg}}$ to $R_{\text{eff}}$. To derive a practical design guideline, we assume that each modulated WDM signal occupies a two-sided bandwidth $B$. The frequency separation between the edges of adjacent WDM channels is $\Delta\nu - B$. The corresponding separation along the $x$-axis on the SLM plane is $((\Delta\nu - B)/\Delta\nu)w_{\text{seg}}$. Obtaining high isolation between adjacent channels requires that this separation not be smaller than the diameter of the beam image on the SLM plane, $2R_{\text{eff}}$. Thus, the minimum switching segment width is

$$w_{\text{seg}} = \frac{\Delta\nu}{\Delta\nu - B} 2R_{\text{eff}}, \qquad (3)$$

where $R_{\text{eff}}$ is obtained from Fig. 3 based on one of the three criteria, and is replaced by $R_{\text{eff},0}$ for SMF.

Expression (3) describes the fundamental scaling principle that the switching segment width $w_{\text{seg}}$ relative to the image size of $R_{\text{eff}}$ should remain constant. Hence, in converting from a single- to a multi-mode beam, the switching segment width $w_{\text{seg}}$ relative to the single-mode image size $R_{\text{eff},0}$ should be scaled by $\kappa$. This scaling principle is satisfied by all the designs in Table I. According to (3), the switching segment width $w_{\text{seg}}$ must be increased further in inverse proportion to $\Delta\nu - B$.

In an LCoS-based WSS, the ruled grating geometry and any additional anamorphic optics transform the image of a beam on the LCoS SLM into an elliptical spot. The image spot is compressed along the frequency direction ($x$-axis) in order to accommodate many WDM channels in an SLM of limited dimensions. We assume that in all designs, the image eccentricity remains approximately unchanged from a single- to a multi-mode WSS.

The SLM is used to apply a linear phase ramp to an optical beam to steer it to different output ports. For a given maximum steering angle, in a single-mode WSS, the beam-steering ability is determined by number of pixels within the fundamental mode radius $w_0$ or the corresponding $R_{\text{eff},0}$. In a multimode WSS, assuming number of pixels within $R_{\text{eff},0}$ remains constant, the number of pixels of SLM along the beam-steering direction ($y$-axis) must be scaled by $\kappa$ to maintain the same beam-steering performance for the higher-order modes. In Designs I-III, the number of pixels within fundamental mode radius $w_0$ remains the same. In Designs I and III, which use the same SLM pixel pitch, the dimensions of the SLM along both the $x$ and $y$ directions must be scaled by a factor of $\kappa$.

In Design I, the ruled grating angular dispersion $\partial\theta/\partial\nu$ remains unchanged from a single- to multi-mode WSS. From (2), in order to increase $w_{\text{seg}}$ by a factor $\kappa$, the Fourier lens focal length $f_{\text{Fourier}}$ must increase by a factor $\kappa$. The image size of all modes on the ruled grating is determined by the inverse Fourier transform performed by the Fourier lens. The image of the fundamental mode on the ruled grating is a factor $\kappa$ larger in a multimode WSS than in a single-mode WSS due to the increased Fourier lens focal length.

In Design I, to maintain the same image size for the fundamental mode with $f_{\text{Fourier}}$ increased by $\kappa$, the collimator lens focal length $f_{\text{coll}}$ must also scale by $\kappa$, as the magnification is related to the ratio of the two focal lengths, which in turn implies that the fundamental mode becomes a factor of $\kappa$ larger at the collimator lens. However, for a multimode beam, the effective beam radius $R_{\text{eff}}$ is itself a factor $\kappa$ larger, so the radius of the collimator lens must be a factor $\kappa^2$ larger than that in a single-mode WSS. The minimum spacing between two ports is mainly determined by the collimator lens radius, so the port spacing must increase by a factor $\kappa^2$.

Similarly, the ruled grating dimensions along both directions must increase by a factor $\kappa^2$. Also, the radius of the Fourier lens should scale by a factor $\kappa^2$, assuming a fixed number of ports, so its $f$-number should scale by a factor $1/\kappa$, which may become problematic for large $\kappa$ if the $f$-number becomes impractically low.

An additional factor considered here is that in going from SMF to MMF, the fundamental mode radius $w_0$ is scaled by a factor $\eta$ at the fiber input and output facets. Unlike the scaling associated with an increase in the number of modes, in which both the beam radius and NA are scaled by $\kappa$, when the fundamental mode radius $w_0$ changes by a factor $\eta$, the NA of the fundamental mode changes by a factor $1/\eta$. The NA of the MMF is $\kappa_{\text{NA}}/\eta$. A change in $w_0$ can be accommodated by modifying the collimator lens focal length and radius to keep the collimated fundamental-mode beam radius constant.

Under Design I, the collimator lens should magnify the beam to a radius $\kappa^2$ larger. Assuming the fundamental mode radius $w_0$ scales by a factor of $\eta$ from SMF to MMF, the collimator lens focal length $f_{\text{coll}}$ should also scale by an additional factor of $\eta$ in order to maintain the same spot size at the SLM. However, because the measured NA of the fundamental mode scales changes by a factor of $1/\eta$ from SMF to MMF, the collimator lens radius is independent of $\eta$. For Design I, the overall scaling factor for the collimator lens focal length is $\eta\kappa$, and its $f$-number should be scaled by a factor $\eta/\kappa$.

The port spacing, which is determined by the collimator lens diameter, increases by a factor $\kappa^2$. With the increase in the focal length of the Fourier lens, the angular separation between two adjacent ports increases by a factor $\kappa$. Keeping the same SLM pixel pitch, the maximum beam-steering angle remains unchanged, so under Design I, the number of ports is reduced by a factor of $\kappa$ as compared to a SMF WSS.

Design II scales the pixel pitch of the LCoS SLM by a factor $1/\kappa$ from the single-mode WSS design, such that the image of the fundamental mode on the SLM can be scaled by a factor $1/\kappa$ from the single-mode WSS. The number of pixels across the image of the fundamental mode on the SLM along the $x$- and $y$-directions remains constant in scaling from SMF to MMF. All of the scaled component parameters for Design II are listed in Table I. The overall dimensions of the LCoS SLM and the Fourier lens focal length $f_{\text{Fourier}}$ are scaled by a factor of $1/\kappa$ compared to Design I, and are the same as for the single-mode WSS. But the Fourier lens $f$-number must be scaled by a factor of $1/\kappa^2$ from the single-mode WSS, which is more problematic than the scaling in Design I for large $\kappa$. The number of ports is also scaled by a factor of $1/\kappa^2$ as compared to the single-mode WSS.

Design III scales up the ruled grating angular dispersion $\partial\theta/\partial\nu$ by a factor of $\kappa$ from the single-mode WSS design, so the switching segment width $w_{\text{seg}}$ may scale by a factor $\kappa$ without changing the Fourier lens focal length $f_{\text{Fourier}}$, as seen from (2). All of the scaled component parameters for Design III are listed in Table I. Design III reduces many key dimensions within the WSS by a factor $1/\kappa$ as compared to Design I, including the ruled grating size, Fourier lens focal length and radius, collimator lens radius, and port spacing. However, increasing the ruled grating angular dispersion can be problematic, especially if it has been highly optimized in the initial single-mode WSS design.

In Designs I-III, the number of ports is reduced as compared to a SMF WSS. To maintain the same number of ports, the maximum beam-steering angle of the SLM needs to be increased. The beam-steering angle is proportional to the slope of the linear phase ramp applied to the SLM. When a linear ramp is approximated by a stair-step function, the accuracy of the approximation is determined by the product of the step width (the pixel pitch) and the step height (which is proportional to the slope of the linear ramp or the beam-steering angle). Maintaining the same phase accuracy, a factor of $\kappa$ increase in maximum beam-steering angle can be achieved by scaling the pixel pitch by a factor of $1/\kappa$. Design IV is the same as Design I but with the SLM pixel pitch scaled by a factor of $1/\kappa$ (as in Design II), so the number of ports can be the same as that of the SMF WSS.

Other design choices may combine various aspects of the designs shown in Table I, for example, increasing the ruled grating angular dispersion by a factor $\kappa$ and reducing the SLM pixel pitch by a factor $1/\kappa$ to obtain the same performance as Design IV.

The analysis given in this section describes a scaling from single- to multi-mode operation based solely on the increased effective beam radius $R_{\text{eff}}$. Unfortunately, a beam of radius $R_{\text{eff}}$ cannot be precisely related to a rectangular SLM segment of width $w_{\text{seg}}$. For example, satisfying (3) with $R_{\text{eff}}$ equal to the 95% beam radius in Fig. 3(b) does not imply a 5% power loss at the SLM, even for the pure modes defining $R_{\text{eff}}$ in Fig. 3(b) (the actual power loss is 1.4%). Because of the complexity of modal profiles (see Fig. 2), different modes with the same $R_{\text{eff}}$ may be subject to different passband shapes. An analysis more precise than (3) is required, and is the subject of the following sections.

### III. FILTERING AND MODE COUPLING FOR PURE MODES

Because of the complex profiles of higher-order pure or mixed modes, evaluating WSS performance requires analysis more detailed than that in Section II. In this section, we discuss filtering of pure modes, while in the following section, we discuss filtering of mixed modes.

#### A. Transmission Coefficients and Mode-Clipping Analysis

Numerical simulations of physical optics propagation have been performed in Zemax for a WSS like that shown in Fig. 1. The channel spacing is $\Delta\nu = 50$ GHz. The starting point is a single-mode WSS that achieves a one-sided 0.5-dB (94.4% transmission magnitude) bandwidth of about 22.0 GHz [35]. Using the scaling of Design II in Table I, a multimode WSS is designed for five mode groups ($g_{\max} = 5$), a total of $D = 30$ modes in two polarizations. A scaling factor $\kappa = 2$ is used. Figs. 3(a) and (b) suggest that $\kappa \approx 1.8$ and 1.9 based on NA and 95% beam radius criteria, respectively, would suffice.

The transmission characteristics of a multimode WSS are somewhat more complicated to characterize than those of a single-mode WSS. One may extend the conventional single-mode power transmission coefficient by computing the power transmission from a specific input mode to all propagating modes in the output fiber. Here, in order to be able to study mode-coupling effects, we compute the amplitude transmission coefficient from a specific input mode to a specific output mode.

Figure 4 shows the magnitudes of the frequency-dependent amplitude transmission coefficients of the multimode WSS for selected modes, where a specific mode at the input is coupled to an identical mode at the output. Figure 4(a) is for the cosine modes shown in Fig. 2 and Fig. 4(b) is for the corresponding sine modes. Figure 4 shows $LG_{00}$ and $LG_{01}$, which are the modes in the two lowest groups, as well as $LG_{12}$ and $LG_{04}$, which are the modes in the highest group with $g = 5$ with both sine and cosine modes. The $LG_{12}$ mode has a size close to that of the largest $LG_{20}$ mode in Fig. 3, but has a more complex mode structure.

Due to the symmetry of the pure modes, the transmission coefficients are symmetric with respect to the center of the WDM channel, so only the positive-frequency side is shown in Fig. 4. For all these pure modes, the one-sided 6-dB (50%) bandwidth is very close to $\Delta\nu/2 = 25$ GHz. The worst-case (minimum) one-sided 0.5-dB bandwidth is about 20.4 GHz. The worst-case one-sided 3-dB (70.7%) bandwidth is 22.7 GHz. $LG_{00}$ and $LG_{01}$-cosine modes have the same transmission coefficients. $LG_{00}$ and $LG_{01}$-sine modes have the same spatial variation along the $x$-axis, leading to the same transmission coefficients. Similarly, the $LG_{12}$-sine and $LG_{04}$-sine modes have the same transmission coefficients.

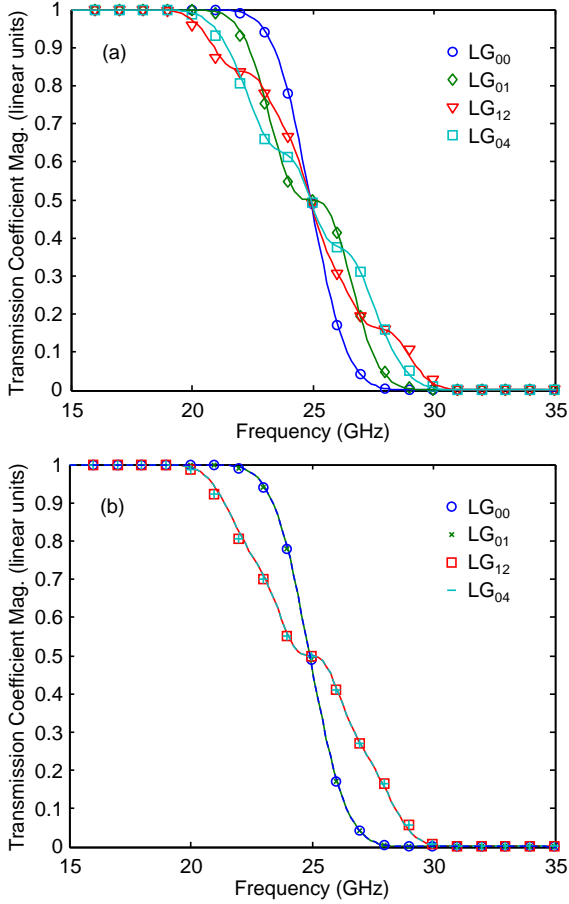

Fig. 4. Transmission coefficients of a multimode WSS for selected (a) cosine modes and (b) sine modes. Results from simulation and the clipping model are shown as symbols and curves, respectively

In Fig. 4, these simulation results are compared with a mode-clipping model in which all light outside the nominal SLM switching segment is assumed lost. In the clipping model, the frequency-dependent amplitude transmission coefficient for mode $(q, m)$ is given by

$$t_{q,m}(f) = \int_{-w_{seg}/2}^{+w_{seg}/2} \int_{-\infty}^{+\infty} \left| E_{q,m}\left(\sqrt{y^2 + [x-l(f)]^2}, \phi\right) \right|^2 dy dx, \quad (4)$$

where $l(f)$ is the center of the beam at frequency $f$ and is a linear function of $f$. When a pure mode is input to the WSS, the coefficient (4) yields the coupling back to the same pure mode, similar to the model in [36] for SMF. Energy may also couple to other modes, as discussed below. Integrating the power over the whole switching segment, expression (4) is also the maximum possible power that can be coupled back to the MMF, with equality only for a MMF with an infinite number of modes.

In (4), the $x$-axis corresponds to the angle $\phi = 0$. The origin in (4) is the center of the switching segment. Although the beam on the SLM surface is elliptical, only the spatial variation along the frequency direction ($x$-axis) affects the filtering response.

Figure 4 shows that transmission coefficients computed by the mode-clipping model (4) are consistent with simulation results. On the linear scale of Fig. 4, the difference appears very small. The difference between the simulation and theoretical results is less than 0.3 dB for normalized losses smaller than 12 dB. Although the difference measured in dB increases at frequencies far from the center of the passband, the absolute transmission and the absolute error are both very small at those frequencies. Figure 4 demonstrates that the mode-clipping model can be used to accurately compute the transmission characteristics of multimode WSSs for LG modes. The LG modes given by (1) are an approximation to the exact modes of a weakly guiding finite-core graded-index MMF, with discrepancies increasing for higher-order modes. It is possible that the mode-dependent transmission coefficients for the higher-order exact modes at the passband edge may be slightly different from those shown in Fig. 4.

Figure 4 shows that the 0.5-dB bandwidth decreases with increasing mode group, due to an increase in beam radius. For all the modes of Fig. 4, and for all the 15 spatial modes within the first five groups, the narrowest 0.5-dB bandwidth is about 20.4 GHz, as shown by both the mode-clipping model and simulation.

The coefficient (4) assumes that only the SLM segment clips the mode, since other components should have a smaller effect. Using the 95% or 99% radius $R_{eff}$ defined in Figs. 3(b) or (c) to compute a scaling parameter $\kappa_{95}$ or $\kappa_{99}$ does not imply that all the components used in the system are chosen to have that radius. Instead, it implies that all the components are scaled according to changes in that radius. Nevertheless, if the lens radius were chosen to equal the 95% beam radius $R_{eff}$ in Fig. 3(b), the transmission coefficient (4) would be reduced at most by 5% with respect to its peak value, with the reduction becoming smaller near the passband edge.

*B. Mode-Coupling Coefficients*

At each frequency, the mode-clipping model (4) can be generalized to compute a frequency-dependent coupling coefficient from mode $(q, m)$ to mode $(p, n)$:

$$c_{q,m;p,n}(f) = \int_{-w_{seg}/2}^{+w_{seg}/2} \int_{-\infty}^{+\infty} E_{q,m}(f,\phi) E_{p,n}^*(f,\phi) dy dx \quad (5)$$

with

$$E_{q,m}(f,\phi) = E_{q,m}\left(\sqrt{y^2 + [x-l(f)]^2}, \phi\right)$$

and

$$E_{p,n}(f,\phi) = E_{p,n}\left(\sqrt{y^2 + [x-l(f)]^2}, \phi\right).$$

The coupling coefficients are symmetric, i.e., $c_{q,m;p,n} = c_{p,n;q,m}$. Note that the coupling coefficient (5) reduces to the transmission coefficient (4) for $(q,m) = (p,n)$. The coupling coefficients (5) are the elements of a real, symmetric mode-coupling matrix, which is used to analyze mixed-mode effects below.

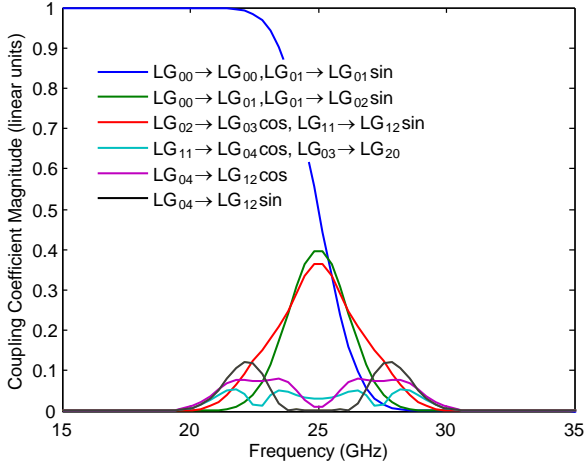

Fig. 5. Coupling coefficients between selected modes for a multimode WSS supporting five mode groups. For comparison, the blue curve shows the transmission coefficients for two modes.

Due to symmetry of the LG modes in (1), sine and cosine modes do not couple to each other in (5). Sine and cosine modes are symmetric and anti-symmetric along the *y*-axis, respectively, and the integration of (5) between cosine and sine modes yields zero. The notation for the mode-coupling coefficient (5) ignores the distinction between cosine and sine modes, with the understanding that sine and cosine modes can be analyzed separately. All modes with zero azimuthal order ($m = 0$) couple only to cosine modes and do not couple to sine modes. Here, all modes with $m = 0$ are classified as cosine modes for convenience.

Figure 5 shows the magnitudes of the coupling coefficients between selected modes for a MMF WSS designed for five mode groups, as in Fig. 4. For comparison, the blue curve shows the transmission coefficients for two modes. In Fig. 5, mode coupling becomes significant only near the passband edge in the frequency range of 20-30 GHz, similar to results in [16][17]. Apart from the symmetry $c_{p,m;q,n} = c_{q,n;p,m}$, some coupling coefficients between different pairs of modes are equal, similar to the equal transmission coefficients for different modes seen in Fig. 4(b). Figure 5 shows that near the passband edge, coupling between modes may be significant, and may become stronger than the coupling between a mode and itself. The mode-dependent coupling in Fig. 5 causes variations in the transmission coefficient at the passband edge, depending on the mixture of modes.

### C. Scaling of Transmission or Coupling Coefficients

The transmission or mode-coupling coefficients computed for one value of κ can be approximately scaled to obtain the coefficients for other values of κ. In both (4) and (5), the only frequency dependence is in $l(f)$, the center of the beam at frequency *f*. The position of $l(f)$ shifts along the *x*-axis linearly with a change of frequency *f*. In the mode-clipping model, assuming a large segment width $w_{\text{seg}} \gg R_{\text{eff}}$, the coefficients (4) and (5) for different values of κ are of the same functional form, but with a scaling dependent on κ.

Consider Design I and assume $w_{\text{seg}} \gg R_{\text{eff}}$ for all values of κ considered. Given $c_{\kappa_1}(f)$, one of the coupling coefficients (5) for a scaling factor $\kappa_1$, the coupling coefficient for a scaling factor $\kappa_2$ is given by

$$c_{\kappa_2}(f) = c_{\kappa_1}\left(\frac{\Delta \nu}{2} - \frac{\kappa_2}{\kappa_1}\left(\frac{\Delta \nu}{2} - f\right)\right). \qquad (6)$$

The transmission coefficients (4) scale in an identical way.

We observe that (3) may be rewritten as $(\Delta \nu - B)/\Delta \nu = 2R_{\text{eff}}/w_{\text{seg}}$. Given a WSS with a segment width $w_{\text{seg}}$ and nominal channel spacing $\Delta \nu$, if the effective beam radius $R_{\text{eff}}$ is increased, the same WSS can be used, provided the signal bandwidth *B* is reduced in order to increase the ratio $(\Delta \nu - B)/\Delta \nu$. The scaling (6) is consistent with the approximation (3) in which the WSS performance is determined by $\Delta \nu - B$. The scaling (6) is valid for the transmission or coupling coefficients (4) or (5), and for functions of those coefficients. Equivalently, (6) shows that given values of $\Delta \nu$ and $w_{\text{seg}} \gg R_{\text{eff}}$, $\Delta \nu - B$ should be scaled inversely proportional to κ to maintain the same the transmission or coupling coefficients.

### IV. FILTERING EFFECTS FOR MIXED MODES

In long-haul MDM systems, because of mode coupling, signals propagate in mixtures of modes. In this section, we analyze filtering of mixed modes using the mode-clipping model of (4) and (5). First, a statistical analysis of mode-averaged filtering effects in the strong-coupling regime is given. Then, worst-case modes with extreme bandwidth or center-frequency shifts are studied.

### A. Mode-Averaged Filtering

A mixed-mode signal including $g_{\max}$ mode groups may be described as

$$E_{\text{mix}} = \sum_{2q+m+1 \leq g_{\max}} \alpha_{q,m} E_{q,m}, \qquad (7)$$

where the $\alpha_{q,m}$ are the complex modal amplitudes and $E_{q,m}$ are the eigenmodes given by (1). The notation of (7) does not explicitly separate cosine and sine modes, but both are considered in the analysis. A normalization $\sum_{2q+m+1 \leq g_{\max}} |\alpha_{q,m}|^2 = 1$ is assumed.

In system simulation with input signals that are generally time-dependent mixed modes of the form (7), the frequency-dependent mode-coupling coefficients (5) may be used to find the corresponding time-dependent output mixed modes. Such time-dependent simulation may generally be used to find the time-dependent distortion induced by a WSS. In a system with multiple cascaded WSSs, however, this method may become computationally intensive, and the following statistical analysis may be useful.

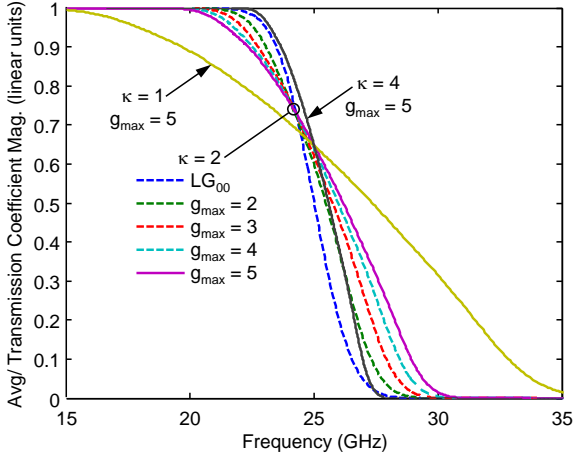

Fig. 6. Mode-averaged transmission coefficients of a multimode WSS for different numbers of mode groups with scaling parameter $\kappa = 2$ and for five mode groups with $\kappa = 1$ and $\kappa = 4$.

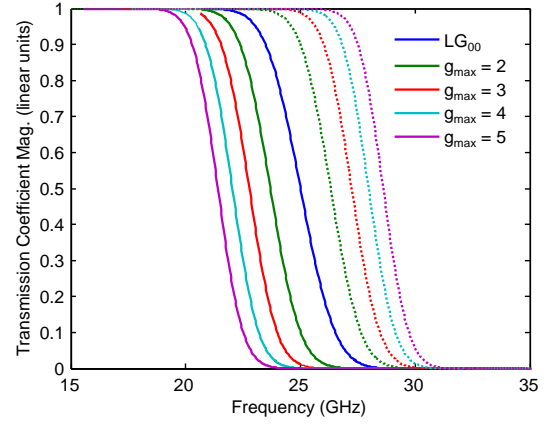

Fig. 7. Transmission coefficients for mixed beams with the narrowest bandwidths (solid curves) or widest bandwidths (dashed curves) for different numbers of mode groups, compared to the $LG_{00}$ mode. Curves with the same color represent the same number of mode groups.

In a long-haul system, a mixed-mode signal (7) may pass through a cascade of many WSSs. The analysis of filtering by the cascade is simplified by considering a large number of WSSs in the strong-coupling regime, which assumes full random coupling of all propagating modes between each WSS. Strong mode coupling is desirable in practice because it minimizes the impact of mode-dependent gain and loss [7][28], and also minimizes MIMO signal processing complexity [27][28].

We wish to compute the transmission coefficient magnitude of the cascade averaged over the ensemble of complex modal amplitudes appearing in (7). In the strong-coupling regime, the complex amplitudes of the eigenmodes in (7) at the input of each WSS are independent and identically distributed, i.e., they have statistically equal powers and random phases uniformly distributed on $[0, 2\pi)$. The ensemble-average correlation between amplitudes is $\langle \alpha_{q,m} \alpha^*_{p,n} \rangle = \delta_{q,p} \delta_{m,n} / D$, where $\langle \, \rangle$ denotes ensemble average. Because $\langle \alpha_{q,m} \alpha^*_{p,n} \rangle = 0$ if $p \neq q$ or $m \neq n$, interference between different eigenmodes does not contribute to mode-averaged filtering effects (but does affect filtering of individual random realizations). Using simple algebra, we find that the average transmission coefficient magnitude per WSS in the cascade is equal to the mode-averaged transmission coefficient magnitude of one WSS for $g_{max}$ mode groups, given by

$$\bar{t}_{g_{max}}(f) = \sqrt{\frac{1}{D} \sum_{2q+m+1 \leq g_{max}} \sum_{2p+n+1 \leq g_{max}} \left| c_{q,m;p,n}(f) \right|^2}, \quad (8)$$

which includes diagonal and off-diagonal elements of the mode-coupling matrix described by (5).

Figure 6 shows the equal-weight mode-averaged transmission coefficient magnitude (8) as a function of frequency. In Fig. 6, mode-averaged transmission coefficients $\bar{t}_5(f)$ for five mode groups ($g_{max} = 5$) are shown for scaling parameter values $\kappa = 1, 2,$ and $4$. The choice $\kappa = 2$ is equivalent to that in Figs. 4 and 5. The choice $\kappa = 1$ (using a single-mode WSS without scaling) causes the passband shape to be degraded significantly, and would lead to substantial bandwidth narrowing, as well as substantial interference from adjacent channels. Conversely, the choice $\kappa = 4$ makes the passband shape more nearly ideal than for $\kappa = 2$. The scaling for different $\kappa$ is consistent with both (3) and (6).

Also shown in Fig. 6 are mode-averaged transmission coefficients (8) for the fundamental $LG_{00}$ mode and for two to five mode groups ($2 \leq g_{max} \leq 5$), for $\kappa = 2$. The passband shape is continuously degraded as the number of mode groups increases, also consistent with (3) with $R_{eff}$ taken from Figs. 3.

### B. Minimum- or Maximum-Bandwidth Modes

The following three sections discuss worst-case mixed modes, which may be useful in conservative system design.

The mixed modes with the narrowest or widest bandwidths can be computed using a matrix whose elements are given by the mode-coupling coefficients (5). As illustrated in Fig. 5, the coefficients (5) are frequency-dependent. The extreme-bandwidth mixed modes are of the form (7) with frequency-dependent modal amplitudes $\alpha_{q,m}(f)$. The narrowest-bandwidth mode has $\alpha_{q,m}(f)$ given at each frequency $f$ by the eigenvector of the mode-coupling matrix (5) that has the smallest eigenvalue, corresponding to the minimum transmission coefficient. Similarly, the widest-bandwidth mode has $\alpha_{q,m}(f)$ given at each $f$ by the eigenvector of (5) that has the largest eigenvalue, corresponding to the maximum transmission coefficient.

Figure 7 shows the magnitudes of the transmission coefficients of the minimum- and maximum-bandwidth mixed modes for a WSS designed for five mode groups with scaling parameter $\kappa = 2$, as in Figs. 4 and 5. The fundamental mode radius $w_0$ is held constant, and the number of mode groups is varied from two to five ($2 \leq g_{max} \leq 5$). The fundamental $LG_{00}$ mode is also shown for comparison. For a given number of mode groups, any transmission coefficient between the minimum and maximum is possible for some set of mixed modes. The transmission coefficients in Fig. 7 are symmetric with respect to zero frequency, so only the positive-frequency side is shown.

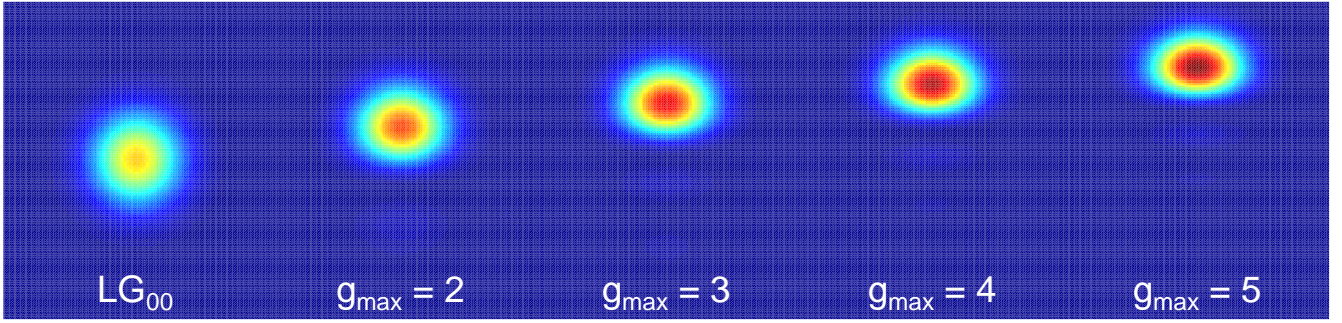

Fig. 8. Intensity patterns at SLM plane for maximum-offset modes for different numbers of mode groups, compared to the $LG_{00}$ mode. The offset increases with an increasing number of groups.

In Fig. 7, including five mode groups, the narrowest and widest 6-dB (50%) bandwidths are 21.3 and 28.6 GHz, respectively. Any 6-dB bandwidth in the range $25.0 \pm 3.6$ GHz is possible. The narrowest and widest 0.5-dB bandwidths are 20.0 and 27.2 GHz, respectively, varying also in the range of $23.6 \pm 3.6$ GHz. The 3-dB bandwidth lies in the range $24.5 \pm 3.6$ GHz.

The extreme-bandwidth modes in Fig. 7 are all given by combinations of only cosine modes (including $LG_{00}$ mode and other modes with $m = 0$). The six sine modes have zero correlation with those cosine modes, and thus do not contribute to the extreme modes in Fig. 7.

The extreme-bandwidth mixed modes correspond to specific frequency-dependent modal amplitudes $\alpha_{q,m}(f)$. In MDM systems with high-order modal dispersion [7][37][38], data signals have frequency-dependent modal amplitudes, and it is possible, though unlikely, for a data signal to align with an extreme-bandwidth mode at all different frequencies. A signal aligned with the minimum-bandwidth mode would be subject to strong distortion, while a signal aligned with the maximum-bandwidth mode would be subject to strong interference from adjacent channels. Any transmission coefficient between the minimum- and maximum-bandwidth modes is possible, and the ratio between them defines the worst-case or peak-to-peak mode-dependent loss.

C. *Maximum-Offset Modes*

In the frequency-dependent mode-coupling coefficient (5), the coupling coefficient magnitude decreases as the beam center $l(f)$ shifts further from the center of the switching segment along the frequency direction ($x$-axis). If we construct mixed modes whose centroid is maximally shifted along the $x$-axis, those modes will have a smaller bandwidth to one side and a larger bandwidth to the other side, and should have approximately the largest shift in passband center frequency. Unlike the extreme-bandwidth modes in Sec. IV.B, these maximum-offset modes are independent of frequency.

To find the mixed modes with maximum frequency offset, the objective is to find the modal amplitudes $\alpha_{q,m}$ in (7) to maximize the mean offset

$$\bar{x} = \iint x |E_{\text{mix}}|^2 \, dx \, dy, \qquad (9)$$

where the integrations are from negative to positive infinity. The mean offset (9) is a bilinear function of the amplitude coefficients $\alpha_{q,m}$, and is maximized by defining an $x$-correlation matrix and choosing $\alpha_{q,m}$ to be the eigenvector having the largest eigenvalue. The $x$-correlation matrix has elements given by

$$\tilde{c}_{q,m;p,n} = \iint x E_{q,m} E_{p,n}^* \, dx \, dy, \qquad (10)$$

where $2q+m+1 \le g_{\max}$, $2p+n+1 \le g_{\max}$, and the integrations are from negative to positive infinity. The $x$-correlation matrix (10) has only a small number of non-zero elements, which are listed in Table II for cosine modes (including modes with $m = 0$) up to five mode groups ($g_{\max} = 5$). For sine modes with $m \ne 0$, only those mode pairs marked in Table II have non-zero elements, which are of opposite sign from the values in Table II.

Figure 8 shows the intensity profiles on the SLM for the maximum-offset modes for a WSS designed for five mode groups with scaling parameter $\kappa = 2$. As in Figure 7, the fundamental mode radius $w_0$ is held constant, and the number of mode groups is varied from two to five ($2 \le g_{\max} \le 5$). The fundamental $LG_{00}$ mode is shown for comparison.

If only two groups ($LG_{00}$ and $LG_{01}$) are used to form an offset mode, both $LG_{00}$ and $LG_{01}$-sine modes have the same amplitudes, and the maximum mean offset is $\bar{x} = w_0/\sqrt{2}$. If three groups ($LG_{00}$, $LG_{01}$, $LG_{10}$, and $LG_{02}$) are used to form an offset mode, the amplitude ratios for $LG_{00}$, $LG_{01}$, $LG_{10}$, and $LG_{02}$ modes are 1, $\sqrt{3}$, $-1$, and $-1$, and the maximum mean offset is $\bar{x} = \sqrt{3/2}\,w_0$. The offset increases with the number of mode groups, as shown in Fig. 8. In Fig. 8, the offset modes with four and five mode groups have mean offsets of $\bar{x} = \sqrt{3/2 + \sqrt{3/2}}\,w_0 = 1.65 w_0$ and $\bar{x} = \sqrt{5/2 + \sqrt{5/2}}\,w_0 = 2.02 w_0$, respectively.

TABLE II. NON-ZERO ELEMENTS OF THE $X$-CORRELATION MATRIX (10) BETWEEN COSINE MODES. THE MODE PAIRS MARKED BY ASTERISKS ALSO HAVE NONZERO CORRELATION BETWEEN SINE MODES.

| $E_{q,m}$ | $E_{p,n}$ | $c_{q,m;p,n} = c_{p,n;q,m}$ |
|---|---|---|
| $LG_{00}$ | $LG_{01}$ | $w_0/\sqrt{2}$ |
| *$LG_{01}$ | *$LG_{02}$ | $-w_0/\sqrt{2}$ |
| $LG_{01}$ | $LG_{10}$ | $-w_0/\sqrt{2}$ |
| *$LG_{02}$ | *$LG_{03}$ | $\sqrt{3}w_0/2$ |
| *$LG_{02}$ | *$LG_{11}$ | $w_0/2$ |
| $LG_{10}$ | $LG_{11}$ | $w_0$ |
| *$LG_{03}$ | *$LG_{04}$ | $-w_0$ |
| *$LG_{03}$ | *$LG_{12}$ | $-w_0/2$ |
| *$LG_{11}$ | *$LG_{12}$ | $-\sqrt{3}w_0/2$ |
| $LG_{11}$ | $LG_{20}$ | $-w_0$ |

Figure 9 shows the magnitudes of the transmission coefficients of the maximum-offset modes of Fig. 8 calculated using the mode-clipping model for two through five mode groups ($2 \leq g_{max} \leq 5$). The transmission for the $LG_{00}$ mode is shown for comparison. The maximum-offset modes have clearly asymmetric passbands, with bandwidths slightly larger than that of the $LG_{00}$ mode. On the negative-frequency side, the 6-dB bandwidths are larger than 25 GHz, increasing to 28.5 GHz for $g_{max} = 5$, potentially increasing interference from an adjacent channel. On the positive-frequency side, the 6-dB bandwidths are smaller than 25 GHz, decreasing to 21.5 GHz for $g_{max} = 5$, potentially increasing signal distortion.

The curves in Fig. 7 with narrowest bandwidth are similar to the positive-frequency side of the curves in Fig. 9. Likewise, the curves in Fig. 7 with widest bandwidth are similar to a folding of the negative-frequency portion of Fig. 9 to positive frequency, corresponding to changing the modes of Fig. 8 from maximum positive to maximum negative offset. The transmission coefficients in Fig. 9 exhibit some differences from those in Fig. 7, particularly ripples on the negative- and positive-frequency sides at large and small transmission coefficient values, respectively. When the 6-dB bandwidths in Fig. 9 are compared to those in Fig. 7, the differences are less than 0.2 GHz (in the worst cases, 21.5 versus 21.3 GHz and 28.5 versus 28.6 GHz). The narrowest 0.5-dB bandwidths in Figs. 9 and 7 are almost identical.

*D. Other Mixed-Mode Characteristics*

Unlike the minimum- or maximum-bandwidth modes found in Sec. IV.B, which are frequency-dependent mixed modes, the maximum-offset modes found in Sec. IV.C using (10) are frequency-independent mixed modes. Similar methods may be used to find frequency-independent mixed modes having other characteristics. For example, the mixed mode having maximum RMS radius is found as an eigenvector of a $\rho^2$-correlation matrix, where $\rho$ is the radial coordinate. Using at least four mode groups, we are able to find mixed modes with RMS radius slightly larger than the maximum RMS radius of the pure modes in those mode groups. These mixed modes only include LG modes that have the same azimuthal order [the same $m$ in (1)], and thus have non-zero $\rho^2$-correlation among them. The passband bandwidth for these mixed modes is typically wider than the narrowest bandwidth among pure modes in Fig. 4, but the transition band, e.g., between the 1- and 10-dB bandwidths, is typically wider.

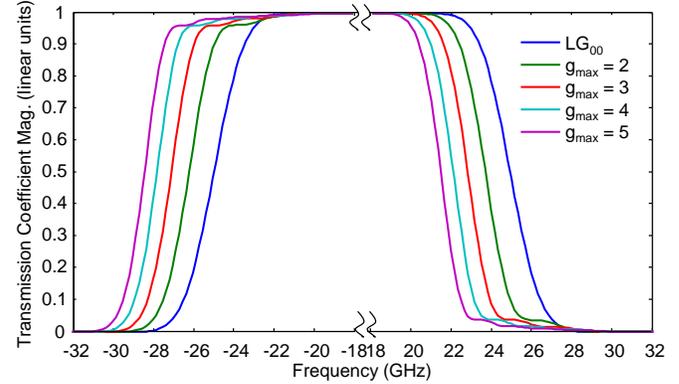

Fig. 9. Asymmetric transmission coefficients of the maximum-offset modes from Fig. 8 for different numbers of mode groups, compared to the $LG_{00}$ mode. The maximum-offset modes have the largest shifts in center frequency.

The mixed modes with the minimum or maximum 0.5-dB two-sided bandwidth may be found approximately as eigenvectors of an $x^2$-correlation matrix (or an $|x|$-correlation matrix) with maximum or minimum eigenvalues, respectively.

V. DISCUSSION

For modeling a cascade of many WSSs with strong mode coupling, the mode-averaged transfer function of Fig. 6 represents a typical response obtained by the law of large numbers. Nevertheless, it is possible to encounter certain mixed modes that are subject to more signal distortion or adjacent-channel interference than the typical case. If the tolerable system outage probability is low, the worst-case mixed modes shown in in Fig. 7 may be used for conservative system design. By adjustment of the scaling parameter $\kappa$, the worst-case minimum bandwidth may be designed to be larger than the signal bandwidth to ensure reliable system performance.

For simulation of time- and frequency-dependent signals in a link, the frequency-dependent WSS mode-coupling coefficients (5) may be used in conjunction with random realizations of fiber propagation matrices to obtain realizations of time- and frequency-dependent output signals. Even if signals occupy a bandwidth more than the minimum bandwidth shown in Fig. 7, simulation results should typically correspond to the mode-averaged transfer function of Fig. 6, since the worst-case mixed modes are unlikely to be encountered, with probabilities of $10^{-6}$ or even lower.

If one frequency-independent mixed mode is required for characterizing WSS performance, especially in experimental measurement, the frequency-independent maximum-offset mixed modes of Fig. 8 may be used to approximate the frequency-dependent worst-case mixed modes of Fig. 7. The difference in 6-dB bandwidth is less than 0.2 GHz.

The coupling coefficients (5), which are important for system performance analysis, may be determined in simulation or measurement by launching into the WSS the inphase and quadrature sum and difference between two pure modes, $E_{q,m} \pm E_{p,n}$ and $E_{q,m} \pm jE_{p,n}$, a total of four combinations. The coupling coefficients can be obtained from the difference between the power transmission coefficients for the sum and the difference between the two modes, e.g., the real part of $E_{q,m}E_{p,n}^*$ is obtained from $\frac{1}{4}\left(\left|E_{q,m}+E_{p,n}\right|^2 - \left|E_{q,m}-E_{p,n}\right|^2\right)$.

## VI. CONCLUSION

WSSs for MDM systems are designed starting with a single-mode WSS and scaling up certain physical dimensions to accommodate the larger size of a multimode beam. All modes at a given wavelength are assumed to be switched as a unit, which is necessary in systems with mode coupling, and minimizes the number of switch ports required to accommodate a given traffic volume.

When a pure mode is present at the switch input, modal transmission coefficients or coupling coefficients are mode-dependent, and may be computed with reasonable accuracy using a simple mode-clipping model. For a given switch design, the bandwidth generally becomes narrower with an increasing number of mode groups.

When multiple modes are present at the switch input, coupling between modes alters the modal transmission and coupling coefficients. Analysis of a system with many cascaded switches with strong mode coupling in between shows that the response of the cascade may be characterized by the mode-averaged transmission coefficient of a single switch. This mode-averaged response incorporates the effect of mode coupling within the switch.

Frequency-dependent mixed modes having the minimum or maximum bandwidth are computed as the frequency-dependent eigenvectors of the mode-coupling matrix with minimum or maximum eigenvalue. The one-sided bandwidth may change up to ± 3.6 GHz. Frequency-independent mixed modes having maximum passband center-frequency shift are computed as eigenvectors of a modal correlation matrix.

## REFERENCES

[1] R. Ryf, S. Randel, A. H. Gnauck, C. Bolle, A. Sierra, S. Mumtaz, M. Esmaeelpour, E. C. Burrows, R.-J. Essiambre, P. J. Winzer, D. W. Peckham, A. H. McCurdy, and R. Lingle, "Mode-division multiplexing over 96 km of few-mode fiber using coherent 6 × 6 MIMO processing," *J. Lightwave Technol.*, vol. 30, no. 4, pp. 521-531, 2012.
[2] D. J. Richardson, J. M. Fini, and L E. Nelson, "Space division multiplexing in optical fibres," *Nature Photonics*, vol. 7, pp. 354-362, 2013.
[3] A. Al Amin, A. Li, S. Chen, X. Chen, G. Gao, and W. Shieh, "Dual-LP$_{11}$ mode 4 × 4 MIMO-OFDM transmission over a two-mode fiber," *Opt. Express*, vol. 19, no. 17, pp. 16672–16678, 2011.
[4] S. Randel, R. Ryf, A. Sierra, P. J. Winzer, A. H. Gnauck, C. Bolle, R.-J. Essiambre, D. W. Peckham, A. McCurdy, and R. Lingle, "6 × 56-Gb/s mode-division multiplexed transmission over 33-km few-mode fiber enabled by 6 × 6 MIMO equalization," *Opt. Express*, vol. 19, no. 17, pp. 16697–16707, 2011.
[5] C. Koebele, M. Salsi, D. Sperti, P. Tran, P. Brindel, H. Margoyan, S. Bigo, A. Boutin, F. Verluise, P. Sillard, M. Bigot-Astruc, L. Provost, F. Cerou, and G. Charlet, "Two mode transmission at 2 × 100 Gb/s, over 40 km-long prototype few-mode fiber, using LCOS based mode multiplexer and demultiplexer," *Opt. Express*, vol. 19, no. 17, pp. 16593–16600, 2011.
[6] P. J. Winzer and G. J. Foschini, "MIMO capacities and outage probabilities in spatially multiplexed optical transport systems," *Opt. Express*, vol. 19, no. 17, pp. 16680–16696, 2011.
[7] K.-P. Ho and J. M. Kahn, "Mode coupling and its impact on spatially multiplexed systems," in *Optical Fiber Telecommunications VI B*, I. P. Kaminow, T. Li and A. E. Willner, Eds., Elsevier, Amsterdam, 2013, ch. 11.
[8] E. B. Basch, R. Egorov, S. Gringer, and S. Elby, "Architectural tradeoffs for reconfigurable dense wavelength-division multiplexing systems," *IEEE J. Sel. Top. Quantum Electron.*, vol. 12, no. 4 pp. 615-626, 2006.
[9] M. D. Feuer, D. C. Kilper, and S. L. Woodward, "ROADMs and their system applications," in *Optical Fiber Telecommunications V*. San Diego: Academic, 2008, ch. 8, pp. 293–343.
[10] S. Tibuleac and M. Filer, "Transmission impairments in DWDM networks with reconfigurable optical add-drop multiplexers," *J. Lightwave Technol.*, vol. 28, no. 4, pp. 557-568, 2010.
[11] T. A. Strasser and J. L. Wagener, "Wavelength-selective switches for ROADM applications," *IEEE J. Sel. Top. Quantum Electron.*, vol. 16, no. 5 pp. 1150-1157, 2010.
[12] J. Homa and K. Bala, "ROADM architectures and their enabling WSS technology," *IEEE Commun. Mag.*, vol. 46, no. 7, pp. 150-154, July 2008.
[13] M. D. Feuer, L. E. Nelson, K. Abedin, X. Zhou, T. F. Taunnay, J. F. Fini, B. Zhu, R. Isaac, R. Harel, G. Cohen, and D. M. Marom, "ROADM system for space division multiplexing with spatial superchannels," in *OFC '13*, paper PDP5B.8.
[14] X. Chen, A. Li, J. Ye, A. Al Amin, and W. Shieh, "Demonstration of few-mode compatible optical add/drop multiplexer for mode-division multiplexed superchannel," *J. Lightwave Technol.*, vol. 31, no. 4, pp. 641-647, 2013.
[15] N. K. Fontaine, R. Ryf, and D. T. Neilson, "Fiber-port-count in wavelength selective switches for space-division multiplexing," *ECOC '13*, paper We.4.B.6.
[16] R. Ryf, N. K. Fontaine, J. Dunayevsky, D. Sinefeld, M. Blau, M. Montoliu, S. Randel, C. Liu, B. Ercan, M. Esmaeelpour, S. Chandrasekhar, A. H. Gnauck, S. G. Leon-Saval, J. Bland-Hawthorn, J. R. Salazar-Gil, Y. Sun, L. Grüner-Nielsen, R. Lingle, and D. M. Marom, "Wavelength-selective switch for few-mode fiber transmission," *ECOC '13*, paper PD1.C.4.
[17] R. Y. Gu, E. Ip, M.-J. Li, Y.-K. Huang, and J. M. Kahn, "Experimental demonstration of a spatial light modulator-based few-mode fiber switch for space-division multiplexing," *FiO '13*, paper FW6B.
[18] J. Carpenter, S. G. Leon-Saval, J. R. Salazar-Gil, J. Bland-Hawthorn, G. Baxter, L. Stewart, S. Frisken, M. A. F. Roelens, B. J. Eggleton, and J. Schröder, "1x11 few-mode fiber wavelength selective switch using photonics lanterns," *Opt. Express*, vol. 22, no. 3, pp. 2216-222, 2014.
[19] L. E. Nelson, M. D. Feuer, K. Abedin, X. Zhou, T. F. Taunay, J. M. Fini, B. Zhu, R. Isaac, R. Harel, G. Cohen, and D. M. Marom, "Spatial superchannel routing in a two-span ROADM system for space division multiplexing," *J. Lightwave Technol.*, vol. 32, no. 4, pp. 783-789, 2014.
[20] N. K. Fontaine, R. Ryf, C. Liu, B. Ercan, J. R. Salaza-Gil, S. G. Leon-Saval, J. Bland-Hawthorn, and D. T. Neilson, "Few-mode fiber wavelength selective switch with spatial-diversity and reduced-steering angle," *OFC '14*, paper Th4A.7.
[21] G. Baxter, S. Frisken, D. Abakoumov, H. Zhou, I. Clarke, A. Bartos, and S. Poole "Highly programmable wavelength selective switch based on liquid crystal on silicon switching elements," *OFC '06*, paper OTuF2.

[22] S. Frisken, I. Clarke, and S. Poole, "Technology and applications of liquid crystal on silicon (LCoS) in telecommunications," in *Optical Fiber Telecommunications VI A*, I. P. Kaminow, T. Li and A. E. Willner, Eds., Elsevier, Amsterdam, 2013, ch. 18.

[23] M. A. F. Roelens, S. Frisken, J. A. Bolger, D. Abakoumov, G. Baxter, S. Poole, and B. J. Eggleton, "Dispersion trimming in a reconfigurable wavelength selective switch," *J. Lightwave Technol.*, vol. 26, no. q, pp. 73-78, 2008.

[24] N. Collings, T. Davey, J. Christmas, D. Chu, and B. Crossland, "The applications and technology of phase-only liquid crystal on silicon devices," *J. Display Technol.*, vol. 7, no. 3, pp. 112–119, 2011.

[25] B. Crosignani, B. Daino, and P. Di Porto, "Speckle-pattern visibility of light transmitted through a multimode optical fiber," *J. Opt. Soc. Am.*, vol. 66, no. 11, pp. 1312-1313, 1976.

[26] O. Wallner, W. R. Leed, and P. J. Winzer, "Minimum length of a single-mode fiber spatial filter," *J. Opt. Soc. Am. A*, vol. 19, no. 12, pp. 2445-2448, 2002.

[27] S. O. Arik, D. Askarov and J. M. Kahn, "Effect of mode coupling on signal processing complexity in mode-division multiplexing", *J. Lightwave Technol.*, vol. 31, no. 13, pp. 423-431, 2013.

[28] K.-P. Ho and J. M. Kahn, "Linear propagation effects in mode-division multiplexing systems," *J. Lightwave Technol.*, vol. 32, no. 4, pp. 614-628, 2014.

[29] E. A. J. Marcatili, "Modes in a sequence of thick astigmatic lens-like focusers," *Bell Sys. Tech. J.*, vol. 43, no. 5, pp. 2887-2904, 1964.

[30] D. Gloge and E. A. J. Marcatilli, "Multimode theory of graded-core fibers," *Bell Sys. Tech. J.*, vol. 52, no. 9, pp. 1563-1578, 1973.

[31] A. E. Siegman, *Lasers*, Sausalito, CA: Univ. Science, 1986, pp. 647-648.

[32] EIA/TIA-455-47B/FOTP-47, Output Far-Field Radiation Pattern Measurement, 1992.

[33] TIA 455-177-A/ FOTP-177, Numerical Aperture Measurement of Graded-Index Optical Fibers, 1992.

[34] W. T. Anderson, "Consistency of measurement methods for the mode field results in a single-mode fiber," *J. Lightwave Technol.*, vol. LT-2, no. 2, pp. 191- 197, 1984.

[35] D. J. F. Barros, J. M. Kahn, J. P. Wilde and T. Abou Zeid, "Bandwidth-scalable long-haul transmission using synchronized colorless transceivers and efficient wavelength-selective switches," *J. Lightwave Technol.*, vol. 30, no. 16, pp. 2646-2660, 2012.

[36] D. M. Marom, D. T. Neilson, D. S. Greywall, C.-S. Pai, N. R. Basavanhally, V. A. Aksyuk, D. O. López, F. Pardo, M. E. Simon, Y. Low, P. Kolodner, and C. A. Bolle, "Wavelength-selective $1 \times K$ switches using free-space optics and MEMS micromirrors: theory, design, and implementation," *J. Lightwave Technol.*, vol. 23, no. 4, pp. 1620-1630, 2005.

[37] M. B. Shemirani and J. M. Kahn, "Higher-order modal dispersion in graded-index multimode fiber," *J. Lightwave Technol.*, vol. 27, no. 23, pp. 5461 - 5468. 2009.

[38] K.-P. Ho and J. M. Kahn, "Frequency diversity in mode-division multiplexing systems," *J. Lightwave Technol.*, vol. 29, no. 24, pp. 3719-3726, 2011.